\documentclass[natbib=true]{sig-alternate-10pt}

\usepackage{grffile} 
\usepackage{xspace}
\usepackage{booktabs}
\usepackage{array}
\usepackage{mdwlist}
\usepackage{url}
\usepackage{subfig}
\usepackage{color}
\usepackage{footnote}
\usepackage{comment}
\usepackage{cite}
\usepackage{balance}

\usepackage{breakurl}
\usepackage[breaklinks]{hyperref}

\usepackage{textcomp}
\newcommand{\mytilde}{{\raise.17ex\hbox{$\scriptstyle\sim$}}}

\graphicspath{{figures/}{graphs/}}
\DeclareGraphicsExtensions{.pdf}

\newcommand{\eg}{\emph{e.g.}\xspace}

\newcommand{\ie}{\emph{i.e.}}
\newcommand{\etc}{\emph{etc.}}
\newcommand{\etal}{{\em et~al.}\xspace}

\newcommand{\rob}[1]{\BREAKME\textbf{\color{red}[ROB: #1]}}
\newcommand{\tim}[1]{\BREAKME\textbf{\color{red}[TIM: #1]}}
\newcommand{\numpages}[1]{}
\newcommand{\checkthis}[1]{{\BREAKME\color{red}#1 {\bf [check this]}}}
\newcommand{\confirmed}[1]{#1}
\newcommand{\fixbreak}{\protect\linebreak}

\newcommand{\ts}{TS\xspace}     
\newcommand{\dc}{DC\xspace}     
\newcommand{\dcs}{DCs\xspace}     
\newcommand{\sk}{SK\xspace}     
\newcommand{\sks}{SKs\xspace}     
\newcommand{\cps}{CPs\xspace}  

\renewcommand{\paragraph}[1]{\vspace{.05in}\noindent\textbf{#1}.~\nolinebreak\nopagebreak}

\hypersetup{
     bookmarks=true,
     colorlinks=true,
     linkcolor=black,
     citecolor=black,
     filecolor=black,
     urlcolor=black,
     pdfauthor = {Akshaya Mani, T Wilson-Brown, Rob Jansen, Aaron Johnson, and Micah Sherr},
     pdftitle = {Understanding Tor Usage with Privacy-Preserving Measurement},
     pdfkeywords = {},
}

\renewcommand{\addauthorsection}{}

\title{Understanding Tor Usage\\with Privacy-Preserving Measurement}

\author{
Akshaya Mani$^{\bullet}$\thanks{Equally credited authors.}
\and
T Wilson-Brown$^{\ddag*}$
\and
Rob Jansen$^{\intercal}$
\and
Aaron Johnson$^{\intercal}$
\and
Micah Sherr$^{\bullet}$
\and
$^{\bullet}$\affaddr{Georgetown University}\\
\affaddr{\{am3227,micah.sherr\}@georgetown.edu}
\and
$^{\ddag}$\affaddr{UNSW Canberra Cyber, University of New South Wales}\\
\affaddr{t.wilson-brown@unsw.edu.au}
\and
$^{\intercal}$\affaddr{U.S. Naval Research Laboratory}\\
\affaddr{\{rob.g.jansen, aaron.m.johnson\}@nrl.navy.mil}
}

\date{}

\begin{document}

\maketitle

\begin{abstract}
The Tor anonymity network is difficult to measure because, if
not done carefully, measurements could risk the privacy (and
potentially the safety) of the network's users.
Recent work has proposed the use of differential privacy and secure
aggregation techniques to \textit{safely} measure Tor, and preliminary
proof-of-concept prototype tools have been developed in order to
demonstrate the utility of these techniques.
In this work, we significantly enhance two such tools---PrivCount and
Private Set-Union Cardinality---in order to support the safe exploration of
new types of Tor usage behavior that have never before been measured.
Using the enhanced tools, we conduct a detailed measurement study of
Tor covering three major aspects of Tor usage: how many users connect
to Tor and from where do they connect, with which destinations do
users most frequently communicate, and how many onion services exist
and how are they used. Our findings include that Tor has \mytilde 8
million daily users (a factor of four more than previously
believed) while Tor user IPs turn over almost twice in a 4 day period.
We also find that \mytilde40\% of the sites accessed over Tor have a
torproject.org domain name, \mytilde10\% of the sites have an
amazon.com domain name, and \mytilde80\% of the sites have a domain
name that is included in the Alexa top 1 million sites list. Finally,
we find that \mytilde90\% of lookups for onion addresses are invalid, and
more than 90\% of attempted connections to onion services fail.
\end{abstract}

\section{Introduction \numpages{1.5}}
The Tor network has been in operation
since \confirmed{2003}~\cite{tor-design}, enabling millions of daily
users~\cite{tormetrics} to anonymously access the Internet.  Despite
its relative longevity, growing popularity, and prominence as a
privacy-enhancing communication tool, surprisingly little
is  known about {\em who} uses Tor and {\em how} they use it.

\paragraph{Challenges in Measuring Tor}
%
The primary challenge in conducting measurements on the Tor network is that, if not done with
extreme care, they can impose significant risk to the network's users.
The {\em collection} of statistical information implies the need to
store data about clients, traffic patterns, sites visited,
{\etc}---information that may otherwise not be collected for privacy reasons.  Even if
the entities collecting this information are honest, the exposure of
the data (\eg, through machine compromise or compulsion from
government authorities) could pose serious privacy risks.
Additionally, the {\em dissemination} of even highly aggregated
measurement information risks users' privacy since adversaries could
potentially combine this data with background knowledge (\eg, when a
tweet was posted, ISP logs showing user accesses to the Tor network,
\etc) to deanonymize users~\cite{dalton-affadavit}.  

In addition to user privacy and safety, monitoring communication is
antithetical to the mission of the Tor network. For this reason, the Tor Metrics
Portal~\cite{tormetrics}, a data repository that archives information
about the anonymity network's size, makeup, and capacity, provides
only a limited number of statistics, many of which are based on
indirect measurements and assumptions of Tor client behavior. We
demonstrate in this paper that some of Tor's estimates do not
correlate well with our own direct observations of user behavior.

Other efforts to measure Tor, such as those that record packets at Tor's
ingress and egress points, can pose serious
ethical and legal issues: they rely on
unsafe~\cite{soghoian-wescr-2011} (and potentially
unlawful~\cite{soghoian-cnet-2008}) techniques that would be difficult
to repeat. As a result, such studies have been met with criticism by the
privacy community~\cite{soghoian-cnet-2008}.

\paragraph{Toward safe Tor measurement}
There has been an exciting recent trend in the literature that
proposes differentially private~\cite{dwork-calibrating-tcc2006}
statistics collection techniques for anonymity networks such as Tor.
Systems such as PrivEx~\cite{PrivExElahi2014},
PrivCount~\cite{privcount-ccs2016}, HisTor$\epsilon$~\cite{histore},
and PSC~\cite{psc-ccs2017} enable useful statistics collection while
providing formal guarantees about the privacy risks they impose on the
network's users.  However, to date, researchers have (rightfully)
focused on developing these privacy-preserving measurement techniques
rather than on performing exhaustive measurements of deployed
anonymity networks such as Tor.


\paragraph{Our Contributions}
This paper presents the largest and most comprehensive measurement
study of the Tor network to date, performed using newly proposed
differentially private statistics collection techniques.  To conduct
our study,
we modify Tor and significantly enhance and improve PrivCount~\cite{privcount-ccs2016} and
PSC~\cite{psc-ccs2017} (described in more detail in the following
section) to support a larger variety of measurements, and contribute our
improvements to the respective open source projects.
We run 16 Tor relays that contribute approximately 3.5\% of Tor's
bandwidth capacity, and deploy our measurement systems across these
relays in order to safely collect measurements of the live Tor
network. We
extend previous statistical methodology to enable unique-counting on
Tor, and extend privacy definitions (“action bounds”) to cover new
types of user activity. We also describe the
challenges and tactics for selecting appropriate privacy parameters to
ensure Tor users' safety, as well as methods for inferring
whole-network statistics given local  observations at our relays. 

Using our improved measurement tools and extended techniques,
we conduct a detailed measurement study of Tor covering three major aspects of Tor usage: who
connects to Tor and from where do they connect; how is Tor used and with
what services and destinations do Tor users communicate; and 
how many onion services exist and how are they used.
Findings from our client-based measurements suggest that Tor has \mytilde 8
million daily users, which is a factor of four more than previously
believed. We are the first to measure client churn in Tor, and we find
that Tor user IPs turn over almost twice in a 4 day period.
We also find that \mytilde40\% of the sites accessed over Tor have a
torproject.org domain name, \mytilde10\% of the sites have an
amazon.com domain name, and \mytilde80\% of the sites have a domain
name that is included in the Alexa top 1 million sites list. Finally,
we find that \mytilde90\% of onion address lookups fail because the 
address is missing or the request is malformed, and
more than 90\% of attempted connections to onion services fail because
the server never completes its side of the connection protocol.
Our findings both
reinforce existing beliefs about the Tor network (\eg, that the vast
majority of its use is for web browsing) and elucidate many aspects of
the Tor network that have previously not been explored.  In some
instances, our measurements---which are based on {\em direct}
observations of behavior on Tor---suggest that existing
heuristically-driven estimates of Tor's usage (including the number of
Tor users) are highly inaccurate.

Limiting the risk to Tor's users was paramount in both the design and
implementation of all of our measurements, and a significant
contribution of this work is the methodology for choosing system
parameters to guarantee adequate levels of protection.  We discuss the
precautions we took to ensure user safety, argue for the ethical
validity of our study, and describe our experiences with institutional
ethics boards and an independent safety review board.

\section{Background \numpages{1}}

We begin by presenting a brief
overview of Tor and reviewing the privacy definitions and
privacy-preserving measurement  tools we use in our
study.  

\subsection{Tor}
\label{sec:background:tor}
Tor provides anonymous TCP connections through source-routed paths
that originate at the Tor client (\ie, the user) and traverse through
(usually) three relays.  Traffic enters the Tor network through {\em
  guard relays}.  Middle relays carry traffic from guard relays to
{\em exit relays}, where the traffic exits the anonymity network.
These anonymous paths through Tor relays are called {\em circuits}.
The unit of transport in Tor circuits is the {\em cell}: Tor cells
carry encrypted routing information and 498 bytes of data~\cite{tor-spec}.
The encryption of application-level headers and payloads makes it
difficult for an adversary to perform traffic analysis on circuits and
learn the endpoints or content of communication.

Tor has a three-tiered communication architecture.  A single circuit
may carry multiple {\em streams}.  A Tor stream is a logical
connection between the client and a single destination (\eg, a webserver),
and is roughly analogous to a standard TCP connection.  A
request to retrieve a webpage may produce several streams (\eg, to
\url{example.com} and  \url{ads.example.com}) that are
multiplexed over a single circuit.
Finally, 
Tor maintains longstanding TLS-encrypted {\em connections}
between relays that are adjacent on some Tor circuit.  Communication
belonging to different circuits that share a common hop between two
relays are multiplexed through a single connection.  Similarly, a
user's Tor client maintains a single connection to each of its guards,
through which multiple circuits (and streams) may be formed.

Tor clients periodically download ``directory updates'' that describe
the consensus view of the network. Clients send all user data through
one guard by default, but directory updates are obtained through three
guards by default. Tor clients avoiding censorship may connect to the
network through \emph{bridges}~\cite{matic2017dissecting}, which are
guards whose identities are
not public and are disseminated by Tor to users requesting them.


\paragraph{Onion services}
Tor  allows services to hide their locations through
the use of {\em onion services}~\cite{onion-services}.  As a special
(but common) case, when the onion
service corresponds to a hidden website, we refer to it
as an {\em onionsite}.

To operate an
onion service, the operator selects six relays to serve as
{\em introduction points}.  It then (1)~constructs an {\em onion
  service descriptor} that contains its public key and the identities
of the chosen introduction points, and (2)~forms Tor circuits to each
of the introduction points.  The descriptor is then stored on six
or eight relays~\cite{tor-rend-spec-v2} (depending on Tor version)
using a distributed hash table (DHT)~\cite{consistent-hashing}. These
relays are called onion service directories (for historical reasons,
they are also called ``hidden service directories''
  or ``HSDirs'').
The DHT is
indexed by the descriptor ID, which is derived from the public key. The
address of the service is a domain with an \url{.onion} suffix, which is
derived from the public key.

When a user wishes to access the onion service, it queries the DHT to
obtain the service's onion descriptor, which includes the identities
of its introduction points. It also chooses a relay to serve as a
{\em rendezvous point} (RP), and constructs a circuit to this RP.
Using the public key in the onion
descriptor, the user encrypts its choice of the RP and sends it via a
Tor circuit to one of the onion service's introduction points.  This
choice is then forwarded to the onion service, using the existing circuit
between the introduction point and the service.  Finally, the onion service
constructs a circuit to the RP, through which it can communicate with the
user.

Importantly, Tor conceals the network locations of both the user and
the onion service.  The communications between the user and the
introduction point, the user and the RP, the onion service and the
introduction point, and the onion service and the RP are {\em all}
carried over Tor circuits.

\subsection{Differential Privacy}
\label{sec:background:dp}
Differential privacy~\cite{dwork-calibrating-tcc2006}
is a privacy definition that
provides strong guarantees about how much is revealed by
answering database queries.
Differential privacy
is defined in terms of pairs of {\em adjacent} databases and guarantees that
query responses cannot be used to distinguish such pairs. Typically, databases
that differ only in a single user's data are considerd adjacent, and then
the definition ensures that query responses don't reveal much about any user.






Differential
privacy has previously been proposed as a means to safely study the
Tor
network~\cite{PrivExElahi2014,privcount-ccs2016,psc-ccs2017,histore}.
In existing work, as in this study, {\em adjacency} is defined by
network activity rather than by user.  For example, in
its study of visits to censored sites, PrivEx defines adjacency as
differing by at most six exit connections per
hour~\cite{PrivExElahi2014}.  Under this definition of adjacency,
differential privacy's guarantees apply to {\em connections} rather
than-{\em users}.  We also adopt this model and consider privacy
protections over particular {\em user activities} (for example, visiting
onion sites) conducted over a fixed length of time rather than on a
per-user basis. The latter is difficult to achieve, since
  conceptually, a single user (\eg, a botnet) could constitute a
  significant fraction of the activity on the network.

Differential privacy is typically achieved by adding noise to query
results.  This noise is added in a controlled manner such that the
result of the query is nearly the same for adjacent databases.  More
formally, a $(\delta,\epsilon)$-differentially private
mechanism~\cite{dwork-ourdata-eurocrypt2006} is an algorithm
$\mathcal{M}$ such that for all adjacent databases $D_1$ and $D_2$ and
all $S \subseteq \text{Range}(\mathcal{M})$,
\[ 
\Pr[\mathcal{M}(D_1) \in S] \leq \exp(\epsilon) \times
\Pr[\mathcal{M}(D_2) \in S] + \delta
\label{eq:dp}
\] 
where $\delta$ and $\epsilon$ are parameters
that can be used to trade off between privacy and accuracy.  

\subsection{PrivCount}
\label{sec:background:privcount}
PrivCount~\cite{privcount-ccs2016} is a distributed measurement system
that provides $(\epsilon,\delta)$-differentially private statistics about the Tor network.

A PrivCount deployment consists of three components: a tally server
(\ts), at least one data collector (\dc), and at least one share
keeper (\sk).  In an execution of PrivCount, the \ts, \dcs, and \sks
act collectively to report (noisy) counts of the events that were
observed during a collection period.  PrivCount supports both single
number queries (\eg, ``how many clients connected during the
collection period?'') and multiple-counter queries (\eg, ``how many
Tor connections went to \{Google, Amazon, Facebook\} during the
collection period?'')  in which the ``bins'' (i.e., counters) are
independent.  PrivCount is shown to provide
$(\epsilon,\delta)$-differential privacy if (1)~at least one \sk is
honest.

\subsection{Private Set-union Cardinality (PSC)}
\label{sec:background:psc}

PrivCount provides a safe method of counting observed events across a
set of Tor relays.  A limitation of PrivCount (and one shared by other
privacy-preserving measurement systems~\cite{histore,PrivExElahi2014})
is that it cannot determine the count of {\em distinct} values among
the \dcs.  For example, while  PrivCount can answer queries such as ``how
many client connections were observed?'', it cannot answer ``how many
{\em unique} clients connected to Tor?''

To answer questions about the count of distinct values, we use the
private set-union cardinality (PSC) protocol introduced by
Fenske \etal~\cite{psc-ccs2017}.  In PSC, each \dc 
$k$ maintains a set of items
$\mathcal{I}_k$.  PSC computes the cardinality of the union of those
itemsets; i.e., $|\bigcup_k \mathcal{I}_k| + \text{noise}$.  PSC
achieves $(\epsilon,\delta)$-differential privacy, and does not expose
any element of any itemset to an adversary.

The participants in PSC include one or more \dcs and one or more {\em
  computation parties} (\cps), the latter of which
 perform a series of noise additions, verifiable
shuffles, re-randomizations, and aggregation to compute the
final cardinality (plus noise) of the union.

\section{Methodology \numpages{3}}
\label{sec:methodology}

Conducting safe measurements on Tor requires careful experimentation
design, configuration, and deployment.  In what follows, we describe
the methodology we employ to collect and analyze measurements of the
Tor network.

\subsection{Deployment}\label{sec:methodology:deploy}
We use several Tor relays and both PrivCount and PSC to conduct Tor
measurements. We first enhanced the PrivCount version of
Tor\footnote{\url{https://github.com/privcount/tor}} so that the PrivCount
events that Tor emits include additional information about connections,
circuits, and streams as necessary to answer our research questions. We also add
new events that report onion service directory usage
information~\cite{tor-rend-spec-v2, tor-rend-spec-v3}. We ran 16 Tor relays with
our enhanced version of Tor (6 exit relays and 11 non-exit relays).

We then enhance PrivCount to support the new information it receives from Tor,
and to support several new counter types as necessary for our study. Most
significantly, we add support for counting set membership using PrivCount
histograms which we use to measure domain distribution
(\S\ref{sec:measurements:exit}), geopolitical distribution in
(\S\ref{sec:measurements:client}), and onion site distribution in
(\S\ref{sec:measurements:onion}). We also significantly extend PSC: we slightly
modify the original PSC design to include a \ts to coordinate the actions of the
\dcs and \cps, we engineer PSC to collect the PrivCount events emitted by our
relays, and we add support for measuring the number of unique domains~(\S\ref{sec:measurements:exit}), clients (\S\ref{sec:measurements:client}), and
onion sites (\S\ref{sec:measurements:onion}). All of our enhancements have been
merged into the PrivCount and PSC open-source projects.

We set up a PrivCount deployment containing 1 TS, 3 SKs, and 16 DCs (one for
each Tor relay), and we set up a PSC deployment containing 1 TS, 3 CPs, and 16
DCs. We use PrivCount and PSC to repeatedly measure Tor in 24 hour periods,
where each period focuses on measuring a small set of statistics. During some
measurement periods, we use only the subset of the DCs and relays that are in a
position to observe the events of interest in order to reduce operator overhead
and reduce the likelihood of failure. Apart from the cases where server(s) was 
temporarily unavailable, the number of CPs/ SKs we use is greater than or 
equal to the number of relay operators in all our measurements.
To maintain our privacy guarantees, PrivCount and PSC measurements 
are never conducted in parallel, and we always enforce at least 24 hours 
of delay between any sequential measurement of distinct statistics.

\subsection{Privacy}
We protect the privacy of individual Tor users using the methodology developed with
PrivCount~\cite{privcount-ccs2016}. This approach applies differential privacy to protect
a certain amount of network activity within a certain length of time. Limiting the
application of differential privacy to bounded amounts of activity is required for producing
accurate results, as otherwise we must protect a hypothetical individual whose activity constitutes
the majority of network activity, which would necessarily yield inaccurate results. Fortunately,
reasonable Tor activity by an individual should for most actions constitute a small fraction of the
total network activity, and so we can provide privacy while providing accurate network measurements.

Our publishing mechanisms formally apply $(\epsilon,\delta)$ differential privacy to the space of
inputs that includes all possible network traces in a given time period. Inputs are considered
``adjacent'' if they differ only in the activity of a single user within a given length of time,
and all differences in that activity stay within a set of \emph{action bounds}. We use 24 hours as
the length of time defining adjacency, and our action bounds are shown in
Table~\ref{table:action_bounds}.
One way to understand the privacy guarantee that results is that, for any two reasonable
sequences of network actions that a user could perform in a 24-hour period, including nothing at
all, they most likely cannot be distinguished based on the statistics published by our mechanisms.

We derive the action bounds by considering reasonable activities that a Tor user might perform
over 24 hours. In this analysis, we consider protecting the privacy of both Tor clients and onion
services. We take into account how certain type of user actions would translate into observable
actions on the Tor network, such as creating new circuits or sending data cells, based on our
understanding of the Tor protocol and software, taking into account features such as caching
and how streams are assigned to circuits. We consider several types of common Tor activities in this
analysis: web browsing with Tor Browser, chat with the Ricochet P2P onion service~\cite{ricochet},
and running a Web server as an onionsite. We compute the amount of network actions that result from
reasonable amounts of these activities and use the maximum to define the action bound. The
activity that provides the maximum value defining each bound is shown in the final column of
Table~\ref{table:action_bounds}.

For example, we choose to protect connecting to 20 domains through
an exit circuit, which would protect browsing two new websites for each of 10 hours per day, as the
other activities (\ie, chat and onionsites) would not create domain connections. This analysis also
allows for additional page loads within the same site, as they would be assigned to the same
circuit and thus would not be measured as a new domain connection. Note that certain observable
actions apply to all Tor activities, such as creating a TCP connection to a guard, and thus have no
defining activity.

We use privacy parameters $\epsilon = 0.3$ and $\delta = 10^{-11}$. The value of $\epsilon$ is
the same one used by Tor to protect its onion-service statistics~\cite{hsstats-torproposal238}.
We choose $\delta$ such that if there are $n$ Tor users, then $\delta/n$
is still small, which ensures that each user is simultaneously
protected~\cite{dwork2014algorithmic}. For example, with $n = 10^{-6}$ then $\delta/n = 10^{-5}$.

\begin{table}[]
\centering
\caption{Action bounds for measurements.}
\label{table:action_bounds}
\small
\begin{tabular}{lll}
\toprule
\textbf{Action} & \textbf{\begin{tabular}[c]{@{}l@{}}Daily\\ bound\end{tabular}} & \textbf{\begin{tabular}[c]{@{}l@{}}Defining\\ activity\end{tabular}} \\
\midrule
Connect to domain & 20 domains & Web \\
\hline
\begin{tabular}[c]{@{}l@{}}Send or receive\\exit data\end{tabular} & 400 MB & Web \\
\hline
\begin{tabular}[c]{@{}l@{}}Connect to Tor\\ from new IP address\end{tabular} & \begin{tabular}[c]{@{}l@{}}1 day: 4 IPs\\2+ days: 3 IPs\end{tabular} & N/A \\
\hline
\begin{tabular}[c]{@{}l@{}}Create TCP\\connection to Tor\end{tabular} & 12 connections & N/A \\
\hline
\begin{tabular}[c]{@{}l@{}}Create circuit\\through entry guard\end{tabular} & 651 circuits & Chat \\
\hline
\begin{tabular}[c]{@{}l@{}}Send or receive\\entry data\end{tabular} & 407 MB & Web \\
\hline
Upload descriptor & 450 uploads & Onionsite \\
\hline
\begin{tabular}[c]{@{}l@{}}Upload descriptor of\\new onion address\end{tabular} & 3 addresses & Onionsite \\
\hline
Fetch descriptor & 30 fetches & Onionsite \\
\hline
\begin{tabular}[c]{@{}l@{}}Create rendezvous\\ connection\end{tabular} & 180 connections & Chat \\
\hline
\begin{tabular}[c]{@{}l@{}}Send or receive\\rendezvous data\end{tabular} & 400 MB & \begin{tabular}[c]{@{}l@{}}Web or\\ onionsite\end{tabular}\\
\bottomrule
\end{tabular}
\end{table}

\subsection{Statistical Analysis}
Our measurement techniques do not give us a complete and exact view of the total amount of
activity in the Tor network. Because we make measurements from a small subset relays, we only
observe a sample of Tor's overall activity. In addition, our privacy-preserving techniques
intentionally include some random errors in the measurement. We use statistical techniques to
overcome these limitations.

The count measurements (\ie, those made using PrivCount) include noise generated according to
a normal distribution with mean zero and known variance. Thus for those measurements we compute
confidence intervals (CIs) that include the true value with 95\% probability. Moreover,
for many values we wish to infer a network-wide total from our sample. We do so by dividing the
reported values (and their CIs) by the fraction of the observations that our measuring relays make.
For example, we measure 32 million streams using relays that comprise 1.5\% of the exit weight
and including added noise with a standard deviation of 3.1 million. We infer
$(3.2\textrm{e}7 \pm 6.2\textrm{e}6) / 0.015 = 2.1\textrm{e}9 \pm 4.1\textrm{e}8$ streams in the
entire network, where the range provides a 95\% confidence interval.

The unique-count measurements (\ie, those made using PSC) include noise generated according to
a binomial distribution with known parameters. In addition, hash-table
collisions (which  can occur within PSC's internal data storage~\cite{psc-ccs2017}) can cause the
measured value to be smaller than the true value. We adjust for these errors by computing 95\%
confidence intervals using an exact algorithm based on dynamic programming.

Extrapolating unique counts from our sample to the entire network can be a challenge. Unlike with
standard counts, we need to know if the same items in our sample are seen elsewhere or not.
For example, when counting unique domains it could be that the domains we count are each
visited very frequently and thus are observed by all relays, which would indicate that our sample
count is the same as the network-wide count. In several cases, we can handle this using other
information we have about the frequency distribution of the observed items. For example, there is
evidence that domains are visited following a power-law
distribution~\cite{krashakov2006universality,adamic2002zipf}. 
We also make some additional
measurements to help determine likely parameters for these distributions. We can then construct
confidence intervals for the network-wide unique counts by considering the probability of our
local counts given different possible values for the overall count. We use Monte-Carlo simulations
to determine these probabilities for more complicated distributions. Note that in some cases
we cannot identify a likely frequency distribution. In these cases, with an observed value of
$x$ and a fraction $p$ of all observations, we simply present the range of likely network-wide
values to be $[x, x/p]$, where the lower end of the range covers the possibility that each
item is frequently and the upper end covers the possibility of infrequent observations per item.

\section{Exit Measurements \numpages{1.5}}
\label{sec:measurements:exit}

In this section, we present results and analyses of exit measurements taken from
relays in a position to observe Tor's egress traffic. 

\subsection{Overview}

Recall from \S\ref{sec:background:tor} that Tor clients build \textit{circuits} through a
sequence of Tor relays over which they multiplex multiple \textit{streams}. Each
stream is associated with a TCP connection between an exit relay and a
user-specified destination and is used to transfer data between that TCP
connection and the client. For each stream, the client specifies a destination
hostname or IP address which the exit relay requires in order to create a TCP
connection on the client's behalf.

To better understand which web domains Tor users frequently visit, we
instrument PrivCount and PSC to count the \textit{initial} streams that are
created on circuits (\ie, the first stream for each circuit) given that a
\textit{hostname} was provided in the stream by the client and the destination
port requested by the client is a \textit{web} port (either 80 or 443). 
We focus on initial streams because the Tor Browser 
uses a new circuit 
for each unique domain shown in the browser address bar. The initial stream
on a circuit will therefore most directly indicate the user's intended
destination, whereas subsequent streams that are created when loading a page to
fetch embedded resources (\eg, images, scripts, \etc) provide less useful
indications of user intent. Second, we focus on streams that provide hostnames
rather than IPv4 or IPv6 addresses because hostnames can be mapped to web
domains much more easily than IP addresses. Finally, we are interested in web
domains and so we do not measure domains on streams that request connection to
destination ports not traditionally associated with web content.

\subsection{Exit Stream Analysis}
\label{sec:measurements:exit:streams}

We first
measure and analyze the breakdown of \fixbreak streams that provide a hostname
and web port, compared to other types of
streams. These measurements provide useful Tor usage information and also
provide context about the overall fraction of Tor exit traffic that our domain
measurements will cover.

\begin{figure}[t]
	\centering
	\vspace{-5mm}
	\subfloat[]{\label{fig:exit-streams:total}\includegraphics[width=0.32\columnwidth]{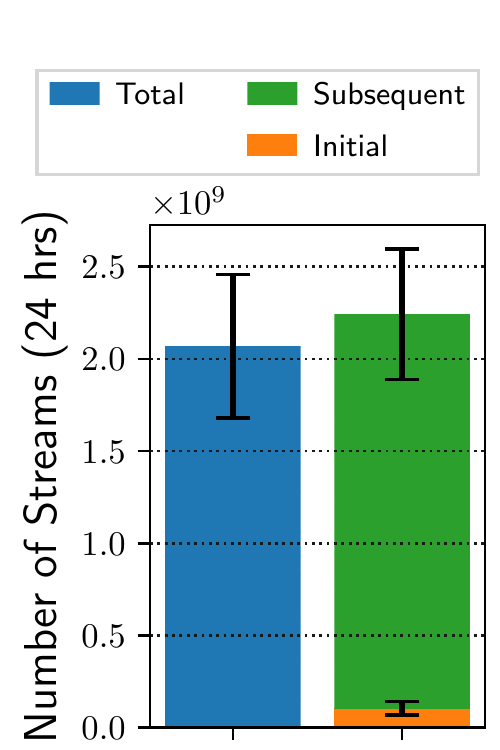}}
	\subfloat[]{\label{fig:exit-streams:initial}\includegraphics[width=0.32\columnwidth]{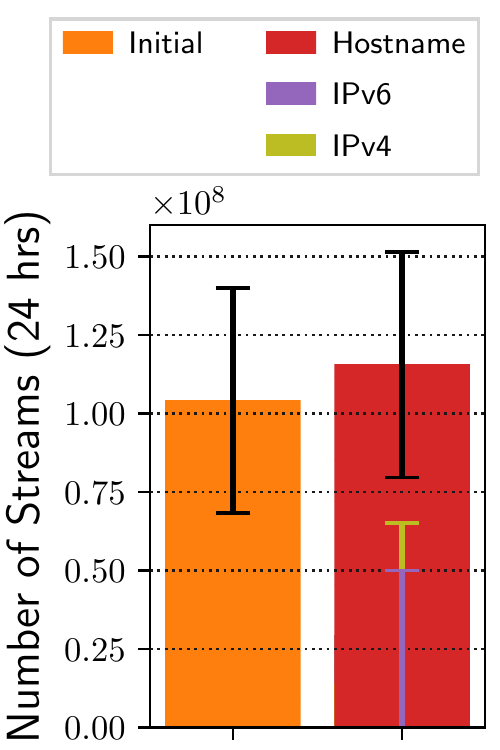}}
	\subfloat[]{\label{fig:exit-streams:hostname}\includegraphics[width=0.32\columnwidth]{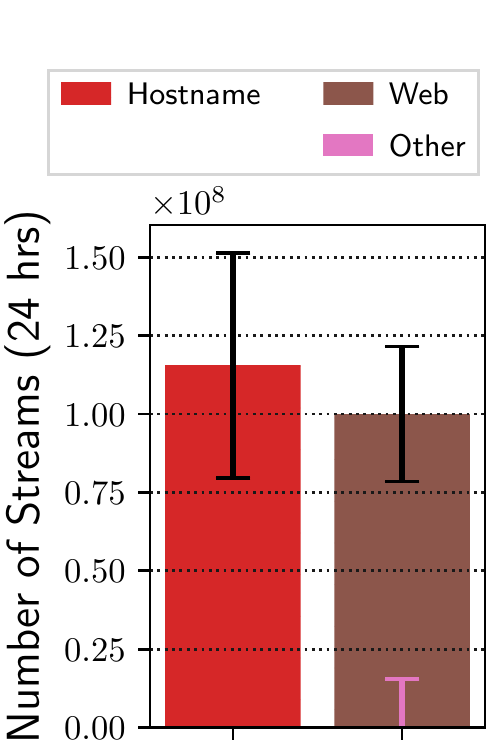}}
	\caption{The number of streams of various types over 24~hours in Tor inferred from our exit relay observations, including 95\% confidence intervals. In each sub-figure, the left bar shows the category total and the right stacked bar shows the breakdown of that category into more specific subcategories.}
	\label{fig:exit-streams}
\end{figure}

Our PrivCount stream measurements were conducted between 2018-01-04 and 2018-01-05 during
which the mean combined exit weight of our measurement relays (taken over the
consensuses that were active during the measurement) was 1.5\% of the total
available exit weight in Tor. We infer the number of streams of various types over a
24~hour period in the entire Tor network using the observations from our relays and
our fractional weight. The inferred values are shown in
Figure~\ref{fig:exit-streams} with 95\% confidence intervals.
Figure~\ref{fig:exit-streams:total} shows that there are about 2 billion exit
streams created in Tor in 24 hours, and only about 5\% of those streams
represent a circuit's first stream. Figure~\ref{fig:exit-streams:initial} shows
that an insignificant number of these initial streams include an IPv4 or IPv6
address: our measured values for these stream subcategories were negative (due
to the added noise) and therefore the most likely value of the counters is zero.
Figure~\ref{fig:exit-streams:hostname} similarly shows that an insignificant
number of initial streams containing a hostname target a non-web port (\ie, a
port other than 80 or 443). Our results confirm that almost all initial streams
provide hostnames and target a web port; the domain measurements that we
describe next will focus on the hostnames provided in these streams.

\subsection{Exit Domains}
\label{sec:measurements:exit:domains}

\looseness-1
We now describe the results from our measurements of the number of domains observed
in initial streams that also provide a hostname and a web port. To ease
presentation, we refer to the measured values as \textit{primary domains}
or \textit{domains} in the remainder of this section.
Across our measurements, we construct sets of domain names and increment a
counter for a set  whenever we observe a primary domain that
matches a domain name in that set.

\paragraph{Alexa Top Sites}
We use the Alexa top 1 millions sites
list\cite{alexa-topsites} to help us
understand which sites are visited by Tor users.  We conducted two
related PrivCount measurements. In the \textit{Alexa rank} measurement, we
sorted the sites by rank and split them into six sets of increasing
size: set $i=0$ contains the first  $10^{1}$ sites and set $i>0$
contains the first $10^{i+1}$ sites excluding those in set $i-1$. We
used a separate set for \url{torproject.org} since early
measurements revealed a significant number of accesses to that domain.
The \textit{Alexa rank} measurement was conducted between 2018-01-31 and
2018-02-01, during which our combined mean exit weight~was~2.2\%.

In the \textit{Alexa siblings} measurement, we created a set for each
of the top 10 sites in the Alexa list. For each such site we stripped
the top level domain to produce a site basename (\eg, google), and
then added all entries from the top 1 million sites list that contained the
basename into the corresponding set. We also used distinct sets for duckduckgo
(rank 342, the default search engine in Tor Browser) and torproject
(rank 10,244, developer of Tor Browser). As a result of this process,
the google set (rank 1) was the largest (212 sites, including the rank
7 site \url{google.co.in}), while the reddit (rank 8) and qq (rank 9)
sets were the smallest among top 10 sites containing 3 sites each (duckduckgo
and torproject contained 1 site each). The \textit{Alexa siblings} measurement
was conducted between 2018-02-01 and 2018-02-02, during which our combined mean
exit weight was 2.1\%.

\begin{figure}[t]
	\centering
	\includegraphics[width=\columnwidth]{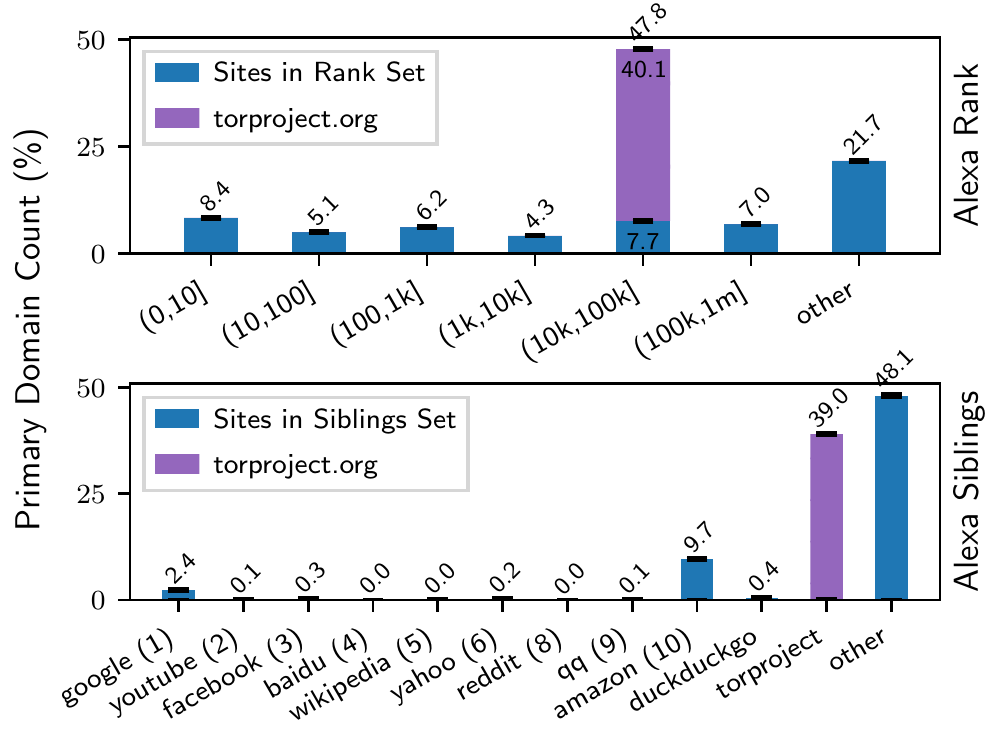}
	\caption{Results of PrivCount measurements of the frequency of membership of primary domains in subsets of the Alexa top
	1 million sites list. The top plot shows subsets containing domains that are sorted by Alexa
	\textit{rank}, and the bottom plot shows subsets containing domains of each of
	the top 10 sites as well as their \textit{siblings}.}
	\label{fig:exit-domains:rank}
\end{figure}

The results from the \textit{Alexa rank} and \textit{Alexa siblings}
measurements are shown in Figure~\ref{fig:exit-domains:rank}. We highlight three
observations from these results. First, we observed \url{torproject.org} in
40.1\% (CI: [39.9;~40.3]\%) 
and 39.0\% (CI: [38.8;~39.2]\%) 
of primary domains. Surprised by this
result, we conducted additional measurements and observed the domain
\url{onionoo.torproject.org} in 
43.4\% (CI: [43.1;~43.7]\%) 
of primary domains. Onionoo
is a web service that provides other applications with access to
Tor network status information. We contacted the Onionoo maintainers from
The Tor Project, but we were unable to identify a clear reason for the high number
of accesses to Onionoo from Tor. 

Second, we observed sites in the amazon siblings set for 
9.7\% (CI: [9.5;~9.9]\%) 
of primary domains, and sites in the google siblings set for 
2.4\% (CI: [2.2;~2.6]\%) 
primary domains. To further explain the unexpectedly high amazon result, we conducted an
additional measurement during which we observed \url{www.amazon.com} in
8.6\% (CI: [8.3;~8.9]\%) 
of primary domains. We contacted Amazon employees but again were
unable to find the reason for the relatively high number of accesses from Tor
(here, due to lack of response).

Third, we observed that domains in the Alexa top sites list accounted
for about 80\% of all primary domains accessed from Tor, and that
the smaller but more highly ranked domain sets have roughly equal
frequency as the increasingly larger but more lowly ranked sets. We
conclude from these results that the Alexa top sites list provides a reasonable
representation of destinations visited by Tor users. The Alexa top sites list is
often used as a destination model in Tor research, and most significantly in Tor
website fingerprinting research~\cite{jansen2018insidejob}. Our results provide
confirmation that using the top sites list as part of Tor research is appropriate.

\paragraph{Alexa Categories}
We conducted a PrivCount measurement of primary domains by category (\eg, news,
science, sports, \etc) using Alexa category
lists~\cite{alexa-categories} which are limited
to 50 sites per category. The measurement was conducted between 2018-01-29 and
2018-01-30, during which our mean combined exit weight was 2.1\%. Additional
insights from the measurement beyond those already presented are limited: the
category containing \url{amazon.com} accounted for 
7.6\% (CI: [7.4;~7.8]\%) 
of primary domains while 
90.6\% (CI: [90.3;~90.9]\%) 
category (\url{torproject.org} was not in any of the categories).

\paragraph{Top-Level Domains}
We measured the frequency with which all top-level domain (TLD) names that are contained in
more than $10^{4}$ entries in the Alexa top 1 million sites list also appeared in the observed
primary domains. The measured TLDs include three main TLDs (\url{.com}, \url{.org}, and
\url{.net}) as well as 11 country-specific TLDs. We conducted two related PrivCount TLD
measurements, one in which we count the TLDs of all primary domains using
wildcards (conducted between 2018-02-02 and 2018-02-03 with 2.4\% exit weight),
and one in which we count the TLDs of only those primary domains that appear in
the Alexa list (conducted between 2018-01-30 and 2018-01-31 with
2.3\% exit weight). We measured \url{torproject.org} using a separate counter
during the Alexa measurement, but our implementation of wildcard matching
restricted us from doing so when measuring all sites.

\begin{figure}[t]
	\centering
	\includegraphics[width=\columnwidth]{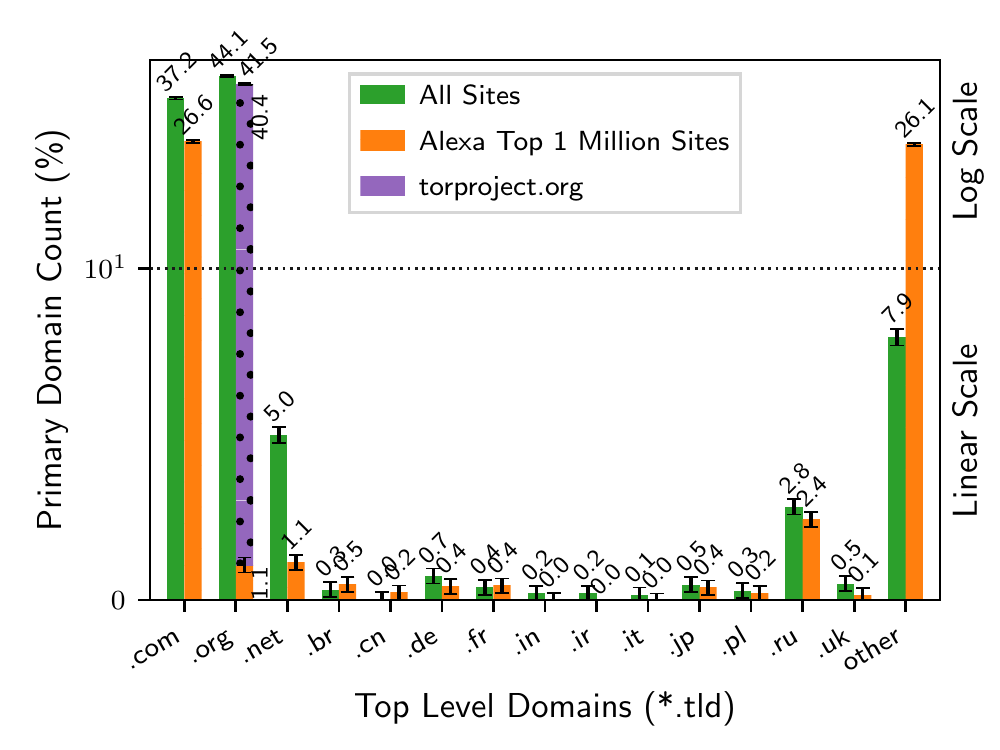}
	\caption{Results of PrivCount measurements of the frequency of membership of primary domains in top-level domain subsets
	constructed from \textit{all sites} and from those sites present in the \textit{Alexa top 1 million sites}
	list.}
	\label{fig:exit-domains:tld}
\end{figure}

The TLD measurement results are shown in Figure~\ref{fig:exit-domains:tld}.
Unsurprisingly, the three main TLDs make up the majority of the primary domains
accessed by Tor users, though we note that \url{.org} would account for a
significantly smaller fraction of domains without \url{torproject.org}. We found
that the Russian TLD \url{.ru} accounted for 
2.8\% (CI: [2.6;~3.0]\%) 
and 2.4\% (CI: [2.2;~2.6]\%) 
of country-specific TLDs, the largest observed by our relays. We also found that
\mytilde11\% of \url{.com}, \mytilde3\% of \url{.org}, and \mytilde4\% of \url{.net} sites
accessed in primary domains are not in the Alexa top sites list. Only 
7.9\% (CI: [7.7;~8.1]\%) 
of TLDs from all primary domains did not match one of the
TLDs that we measured, which increased to 
26.1\% (CI: [25.9;~26.3]\%) 
of TLDs for primary domains in the Alexa top 1 million sites list.

\paragraph{Second-Level Domains}
\begin{table}[t]
  \centering
  \caption{Locally observed {\em unique} second level domain statistics measured using PSC.}
  \label{tbl:PSC:exit:result}
  \small
  \begin{tabular}{l l l}
    \toprule
    {\bf Statistic} & {\bf Count} & {\bf 95\% CI} \\
    \midrule
    SLDs & 471,228 & [470,357; 472,099] \\  
    Alexa SLDs & 35,660 & [34,789; 37,393] \\
    \bottomrule
  \end{tabular}
\end{table}
We conducted two measurements of the number of unique second-level domain (SLD)
names that we observed in Tor primary domains. During both these measurements, 
we use only 5 out of our 6 exit relays (1.24\% mean exit weight) in order to reduce operator 
overhead. We summarize our results in Table~\ref{tbl:PSC:exit:result}.
In the first measurement, we measured the number of unique SLD names among those Tor primary 
domains which contain a TLD in the public suffix list~\cite{pubsuffix}.
From a measurement conducted between 2018-03-31 and 2018-04-01, we find that 
471,228 (CI: [470,357; 472,099]) unique SLDs were accessed through our exits. We also 
measured the number of unique Alexa top 1 million SLDs (i.e. the SLDs of Alexa top 1 million sites) 
among those Tor primary domains. 
From a measurement conducted between 2018-03-24 and 2018-03-25, we infer that
35,660 (CI: [34,789; 37,393]) of such unique SLDs were accessed through our exits. 
From these results, we find that the unique count of  accessed SLDs is
more than ten times that of the unique count of Alexa top one million
sites.  From  our
measurements, we conclude that a long tail exists in the distribution
of sites accessed over Tor.

To extrapolate the unique second-level domain measurements {\em for the entire Tor network}, 
we need to know the distribution of SLDs as seen by the exits. From previous research studies,
we know that domains are visited following a power-law distribution~\cite{krashakov2006universality,
adamic2002zipf}. But we need additional measurements to determine the exponent of this power-law 
distribution. Therefore, we perform a series of simulations of clients visiting random destination sites 
(based on power-law distribution with random exponents) and construct confidence intervals for the 
network-wide unique SLD counts. We use the locally observed unique SLDs count as a self-check. 

This method appears to work well  for the unique Alexa top 1 million
SLDs.  The results of 100 simulations reveals an inferred network-wide
unique count of 513,342  (CI: [512,760; 514,693]) accesses to the
Alexa top 1 million list.  That is, slightly more than half of the
Alexa top sites are accessed over Tor over 24 hours.  Unfortunately,
the inability to closely fit SLD accesses to a distribution prevented
us from using this approach to extrapolate the number of
network-wide SLD accesses.


\section{Client Measurements \numpages{1.5}}
\label{sec:measurements:client}

We conducted a number of measurements to better understand {\em who}
uses Tor and to determine {\em how many} users access the
network.

\subsection{Tor Connections and Clients}
We first measure the number of client connections using our PrivCount
deployment. Over the 24 hour measurement period beginning on
2018-04-07, the mean probability of selecting our relays in the
entry position was 0.0144. Since clients typically choose their guards
according to guard weights, we can project our local PrivCount
measurements to infer Tor-wide results by dividing our observed
counts by 0.0144. We summarize our results in
Table~\ref{tbl:PrivCount:entry:result}.

The amount of daily data being transferred across Tor is 517~TiB (CI: [504; 530]~TiB),
which represents the sum of client uploads and downloads. We note
that this includes Tor cell overheads, so the actual amount of client
payload data transferred would be 2-3\% less.

We find that there are about 148 million (CI: [143; 153] million) client connections,
through which 1,286 million client circuits (CI: [1,246; 1,326] million) are multiplexed, per day,
across the Tor network. This is significantly  higher than the
80.6 million client connections reported by Jansen and
Johnson~\cite{privcount-ccs2016} two years ago. We suspect that the discrepancy is
due to ongoing distributed denial-of-service (DDoS) attacks that began affecting
the Tor network back in December 2017~\cite{tor-ongoing-ddos}.

\paragraph{Tor Clients}
We next  consider the number of  {\em unique} clients access
the Tor network. Here, we use PSC to safely capture the approximate number
of clients observed by our guard relays. Unlike existing work that
also attempts to quantify the number of Tor
users~\cite{chaabane2010digging,privcount-ccs2016,loesing2010measuring,tor-usage},
we do not store (even temporarily) IP addresses since PSC
uses oblivious counters. Similar to previous studies, we
assume that there is a one-to-one mapping
between client IPs and unique Tor clients, although this may be
violated, for example, by mobile users with changing IP
addresses or by clients behind Network Address Translation (NAT). In
addition, we count bridges as clients, as their identities are private.


We use PSC to collect data from our relays that have a non-zero guard
probability, for a 24 hour period beginning on 2018-04-14 during which
our guards had a combined weight of 1.19\%.
Measurement results are reported in Table~\ref{tbl:PSC:entry:result}.
We observe over 313 thousand unique client IPs using our
guards (this experiment used two CPs due to the temporary unavailability of one).
This is a surprisingly high number, given that Tor Metrics~\cite{tormetrics}
(using a very different and unvalidated estimation technique) reports
2.15 million clients per day in April 2018 {\em for the entire Tor network}.
We would expect a typical client to connect
to three guards (clients currently use one guard for data but two additional guards for directory
updates~\cite{tor-dir-spec}), and so we would expect to see closer to
$0.0119 \times 3 \times 2.15\textrm{e}6 = 76,755$ unique clients IPs.
To the extent that each unique IP observed in a 24-hour period represents a new user,
this suggests that Tor is underestimating its total number of users by a factor of 4, and the
total number of daily users is closer to $313,213/0.0119/3 \approx 8,773,473$.

To better understand the network-wide number of unique client IPs in Tor, we perform additional
PSC measurements using sets of relays of \emph{different sizes}. We can then compare how the count
grows with the measuring set to its predicted growth under different client models to identify the
most likely model. In particular we would like to validate a model of how many guards each
client contacts, which we expect to typically be 3, but could be less or more for several reasons,
including guard/exit conflicts, multiple clients behind a single NAT IP, Tor bridges serving many
clients, and tor2web instances~\cite{tor2web}.
We made two 24-hour measurements, one starting on 2018-05-12 using DCs with
0.42\% of the guard weight and the other starting on 2018-05-13 using a disjoint set of DCs
with 0.88\% of the guard weight. We counted 148,174 (CI: [148, 161] thousand) and 269,795
(CI: [269, 315] thousand) unique client IPs during these measurements, respectively. Observe that
the latter number is significantly smaller than we would expect if client contacted only one guard,
in which case we would predict $148,174\times (0.88/0.42) \approx 310,460$. This indicates that
indeed client IPs typically connect to multiple guards.

To identify the number of guards that typical client IPs connect to, we consider a range of guards
per client and analyze the resulting consistency with our measurements. Let $g$ denote the number
of guards each client connects to in the model. We use simulation to
extrapolate from our two measurements two separate confidence intervals for the network-wide
unique client IP counts. We discover that these CIs are \emph{disjoint} unless $g$ is in the range
$[27, 34]$. This is a much higher value of $g$ than should be the case for typical clients, and we
believe that this indicates that this is a poor model of how clients connect to guards. It does
imply, however, that there are some clients IPs that are connected to many more guards than is
typical.

Thus we consider a refinement on this model in which a set of ``promiscuous'' clients connect to
\emph{all} guards in a 24-hour period, and the remaining ``selective'' clients connect to $g$
guards only. The promiscuous clients capture the likely behavior of Tor bridges, tor2web clients,
and clients behind a NAT IP. We then determine a range for the number $p$ of promiscuous clients
by considering values of $g$ that are likely given our understanding of the Tor protocol:
$g\in \{3, 4, 5\}$ (all clients should connect to 3 guards for directory updates, some clients may
connect to 1 or 2 more due to guard churn). For each of these values of $g$, there were some
values for the number $p$ of promiscuous clients that was consistent with our two client-IP
measurements, that is, that yielded a non-empty intersection of the two CIs.
Table~\ref{table:network_client_ips} shows these ranges, which indicates about
14--22 thousand promiscuous clients. Observe that bridges alone seem unlikely to make up this
population, as this range is much larger than the ~1640 bridges reported by Tor Metrics at the time
of our measurements, and in April 2016 Matic et al.~\cite{matic2017dissecting} only discovered 50\%
more bridges than reported by Tor Metrics. Table~\ref{table:network_client_ips} also shows
confidence intervals for the number of network-wide client IPs, calculated for each $g$ as the
union of CIs over all $p$ in the range shown. For what we believe is the most likely number of
guards per selective client ($g=3$), our measurements indicate about 11 million unique client IPs
network-wide, which suggests a true value over 5 times the number of users estimated by Tor Metrics.
Even for $g=5$, the range suggests over 3 times as many daily users as Tor currently estimates.

\begin{table}[t]
\centering
\caption{Network-wide promiscuous clients and client IPs}
\label{table:network_client_ips}
\small
\begin{tabular}{lll}
\toprule
\textbf{\begin{tabular}[c]{@{}l@{}}Guards per\\ client \end{tabular}} & \textbf{\begin{tabular}[c]{@{}l@{}}Promiscuous\\ clients 95\% CI \end{tabular}} & \textbf{\begin{tabular}[c]{@{}l@{}}Network-wide\\ client IPs 95\% CI\end{tabular}} \\
\midrule
3                                                                 & {[}15,856; 21,522{]}                                                  & {[}10,851,783; 11,240,709{]}                                        \\
4                                                                 & {[}15,129; 21,056{]}                                                  & {[}8,195,072; 8,493,863{]}                                          \\
5                                                                 & {[}14,428; 20,451{]}                                                  & {[}6,605,713; 6,849,612{]}\\
\bottomrule
\end{tabular}
\end{table}

\paragraph{Client churn} Tor clients may go offline, and we expect that the Tor network
experiences a high degree of client churn. Using PSC, we recorded the count of unique
client IPs over a four-day period beginning on 2018-05-15. We observed 672,303
(CI: [671,781; 1,118,147]) unique client IPs. Comparing this result with our one-day
measurement (see Table~\ref{tbl:PSC:entry:result}), we can conclude that the client churn
rate is 119,697 (CI: [119,581; 247,268]) client IPs per day. We can infer similar network-wide
behavior \ie client IPs turn over almost twice over a four-day period.

\begin{figure*}[t]
  \centering
  \begin{minipage}[t]{.32\linewidth}
    \centering
    \includegraphics[width=\linewidth]{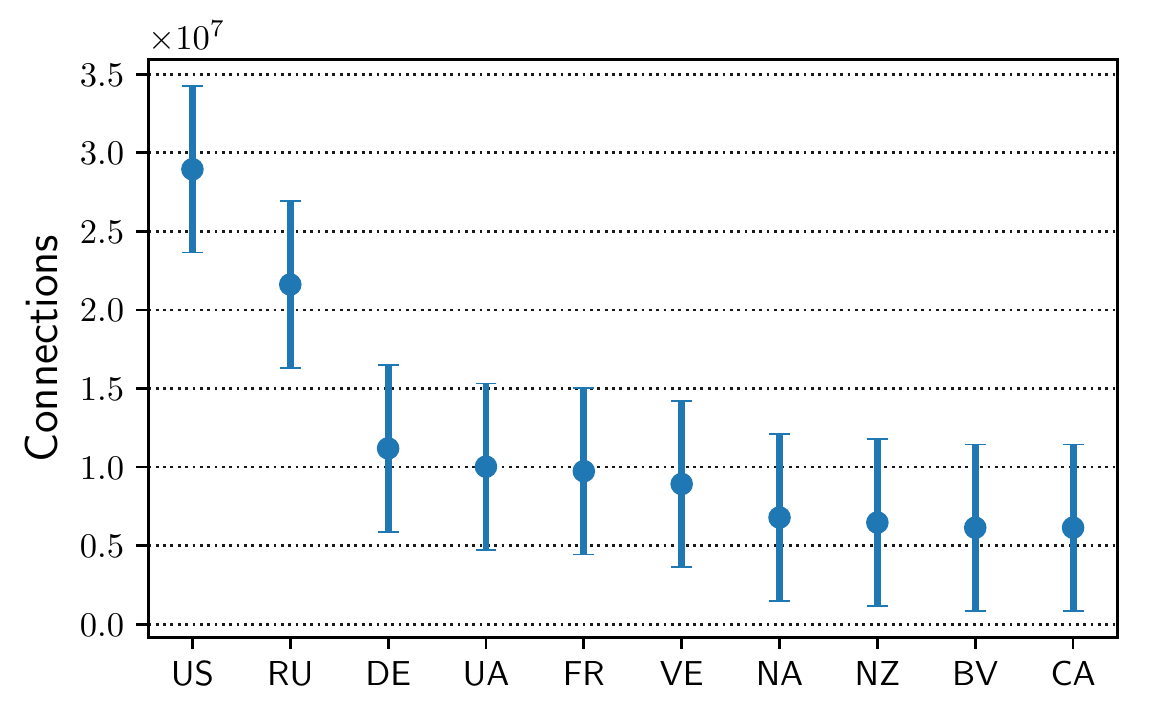}
  \end{minipage}
  \hfill
  \begin{minipage}[t]{.32\linewidth}
    \centering
    \includegraphics[width=\linewidth]{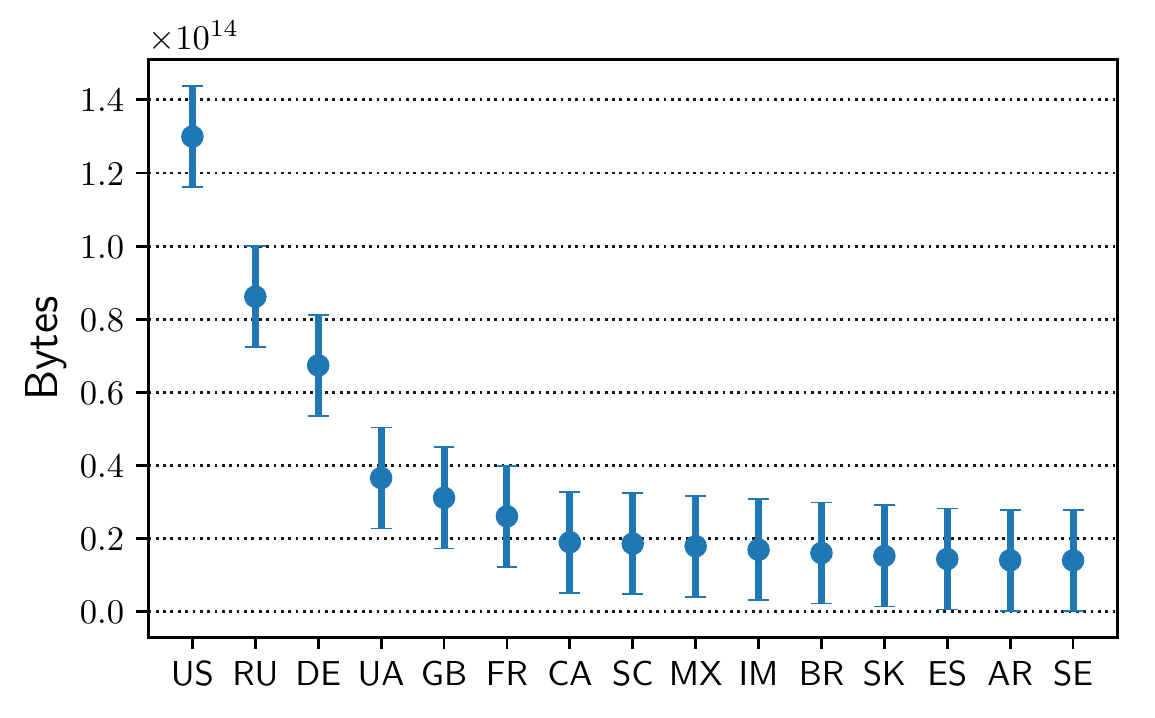}
  \end{minipage}
  \hfill
  \begin{minipage}[t]{.32\linewidth}
    \centering
    \includegraphics[width=\linewidth]{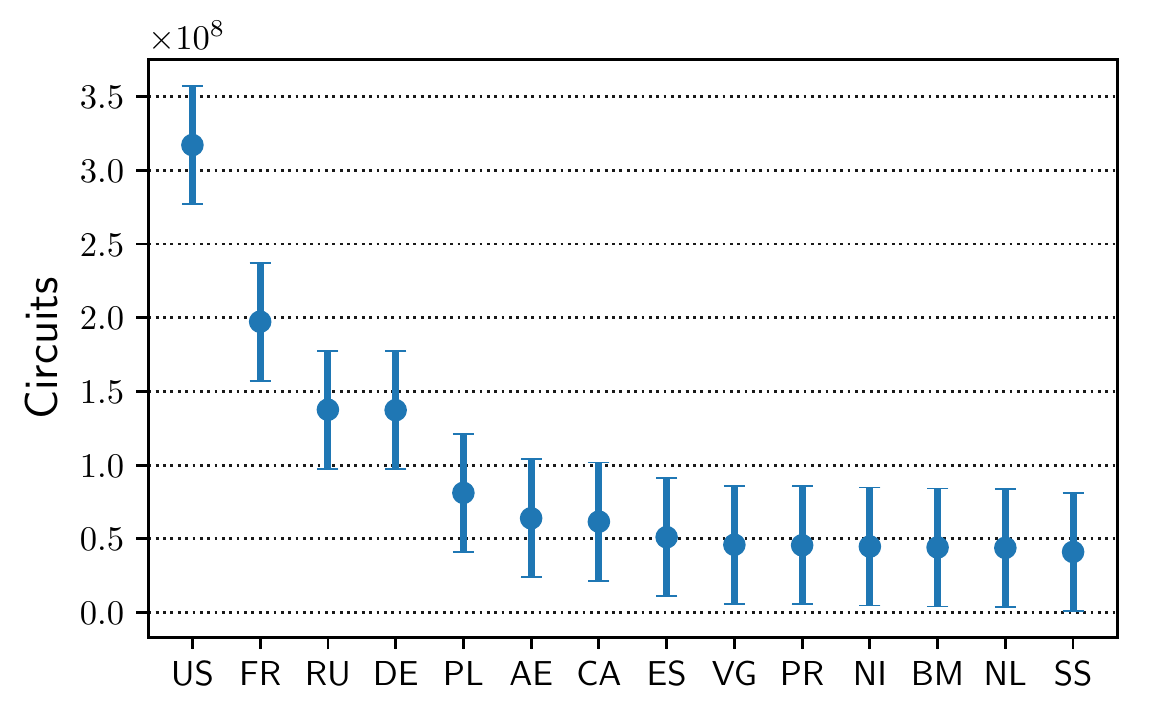}
  \end{minipage}
  \caption{Tor per country client usage statistics inferred from PrivCount measurements -
    the counts of client connections ({\em left}),
    amount of client data transferred ({\em center}),
    and the counts of client circuits ({\em right}).
    Error bars show 95 percent confidence intervals.}
  \label{fig:entry:country}
\end{figure*}

\begin{table}[t]
  \centering
  \caption{Network-wide client usage statistics, inferred from PrivCount measurements.}
  \label{tbl:PrivCount:entry:result}
  \small
  \begin{tabular}{lll}
    \toprule
    {\bf Statistic} & {\bf Count} & {\bf 95\% CI} \\
    \midrule
    Data~(TiB) & 517 &  $\pm$ [504; 530] \\ 
    Connections~($\times10^6$) & 148 & $\pm$ [143; 153] \\
    Circuits~($\times10^6$) & 1,286 & $\pm$ [1,246; 1,326]  \\
    \bottomrule
  \end{tabular}
\end{table}

\begin{table}[t]
  \centering
  \caption{Locally observed {\em unique} client statistics measured using PSC.}
  \label{tbl:PSC:entry:result}
  \small
  \begin{tabular}{l l l}
    \toprule
    {\bf Statistic} & {\bf Count} & {\bf 95\% CI} \\
    \midrule
    IPs & 313,213 & [313,039; 376,343] \\  
    Countries  & 203 & [141; 250] \\
    ASes & 11,882 & [11,708; 12,053] \\ 
    \midrule
    IPs (4-day) & 672,303 & [671,781; 1,118,147] \\
    \quad \quad Churn & 119,697/day & [119,581; 247,268]/day \\
    \bottomrule
  \end{tabular}
\end{table}

\subsection{Client Composition and Diversity}
We additionally explore the geopolitical distribution of Tor clients.
We resolve each client IP address to its host country using the
MaxMind GeoLite2 country database, and use PrivCount to count the
number of client connections, bytes transferred, and circuits created,
broken down by country. For most of the world's 250 countries, we were
unable to determine useful counts since the added noise (to achieve
differential privacy) overwhelm the actual count. We show the
countries with significant inferred counts in
Figure~\ref{fig:entry:country}.

The United States (US), Russia (RU), and Germany (DE) have both the
greatest client connection counts and data transfer amounts (see
Figure~\ref{fig:entry:country}, {\em left} and {\em center}). Assuming
that the distribution of client connection counts reflects the
distribution of client usage, we can surmise that these three
countries use Tor the most.

Interestingly, our results disagree with those from the Tor Metrics
Portal, which ranks the United Arab Emirates (UAE) as contributing the
second largest number of Tor users; in contrast, we did not find the
UAE to be  among the most significant contributors. Although we
cannot fully explain this discrepancy, we are somewhat skeptical of
the Tor Metric Portal's results: 
the value reported by the Tor Metrics Portal implies that nearly 4\%
of the entire population of the UAE accesses Tor on a daily basis,
which seems unlikely. However, although the UAE has
neither a significant client connection count nor client data
transferred, it ranks sixth in the number of client circuits (see
Figure~\ref{fig:entry:country}, {\em right}; marked as ``AE''). One
intriguing possibility is that the majority of the Tor clients from
the UAE are partially blocked from using Tor: while they are able to
construct directory circuits (recall that Tor's user statistics are
estimated from the number of directory requests), they
are prevented from establishing regular Tor circuits, causing the Tor
software to repeatedly perform directory fetches.  We leave the
exploration of this strange Tor client behavior in the UAE as a future
research direction.



To determine whether Tor has a global userbase, we use PSC
to count the number of unique countries from which clients originate.
Since the actual count can only be at most 250 (\ie, the total number
of countries worldwide), invariably the differentially private noise
overwhelms the actual count. Therefore to reduce the effect of noise,
we average the country counts between two consecutive one-day
measurements, beginning on 2018-05-09. We find that clients from
203 (CI: [141; 250]) different countries access Tor. Comparing our
results to earlier studies from McCoy et al.~\cite{tor-usage} (2008)
and Chaabane et al.~\cite{chaabane2010digging} (from 2010) that
report clients from approximately 125 countries, we can conclude
that the locations from which Tor is accessed has significantly
diversified in the past decade.

We were unable to extrapolate the unique country count for the entire
Tor network since the frequency distribution of the countries, as seen
by the guards, is difficult to determine.  However, we can still
roughly estimate the range of network-wide values to be [141, 250], where the
lower bound is the least possible observed count and the upper bound is the
total number of countries that exists worldwide.

\paragraph{Network Diversity}
We also measure the network diversity of Tor users by mapping each
client IP address to its autonomous system (AS) using the IPv4 and
IPv6 datasets (dated 26th November 2017) from
CAIDA~\cite{IPtoAS}.
Using PSC, we first measure the unique AS count of the client IPs, for a
24 hour period beginning on 2018-04-18. We observe clients from about
11,882 ASes (CI: [11,708; 12,053])---approximately 20\% of the number of defined ASes.

To extrapolate the unique AS count {\em for the entire Tor network}, we must first identify
the frequency distribution of the ASes, as seen by the guards. This is challenging.
Therefore, we present the range of most-likely network-wide values as [11,708; 59,597],
where the lower bound is the least possible observed count and the upper bound is the
total number of ASes.

To determine whether there are ``hotspot'' ASes, we also ran PrivCount
and recorded counts for each of CAIDA's 1000 highest ranked
ASes~\cite{asrank}, where an AS's rank is based on the number
of ASes in its customer cone. Over the 24 hour measurement period beginning on 2018-05-01,
we found that none of the Top 1000 ASes had results that were statistically significant (\ie, clearly
distinguishable from noise). Although, we did not find any individual AS that contributes a large
fraction of Tor users, we found that the ASes outside the top 1000 have about 53\% of the client
connections, 52\% of the client data, and 62\% of the client circuits.

\section{Onion-Service Measurements \numpages{1.5}}
\label{sec:measurements:onion}

We conducted a number of measurements to determine {\em how many} Tor
onion services exist, {\em which kinds} of services are popular, and {\em how}
these services are used.

\subsection{Unique Onion Addresses} \label{sec:measurements:onion:address}
\begin{table}[t]
  \centering
  \caption{Network-wide number of \textit{unique} onion addresses inferred from our HSDir PSC measurements.}
  \label{tbl:PSC:onion:result}
{\small
  \begin{tabular}{lll}
    \toprule
    {\bf Statistic} & {\bf Count} & {\bf 95\% CI} \\
    \midrule
    Addresses published & 70,826 & [65,738; 76,350] \\
    Addresses fetched & 74,900 & [34,363; 696,255] \\ 
    \bottomrule
  \end{tabular}
}
\end{table}

Recall from \S\ref{sec:background:tor} that Tor onion services store their
descriptors at onion service directories (HSDirs) in a DHT.
To measure the number of \textit{unique} onion services in
the Tor network, we instrument our HSDirs to
report the onion service \textit{address} for each version 2 (v2)~\cite{tor-rend-spec-v2}
descriptor that is \textit{published} to and \textit{fetched} from the DHT. (We
don't measure version 3 (v3) onion service descriptors
because the onion address is obscured using key blinding~\cite{tor-rend-spec-v3}.)
We use the PSC deployment described in~\S\ref{sec:methodology:deploy} to safely capture the approximate number of v2 onion services
observed by our HSDirs. Unlike existing work
that also attempts to quantify the number of onion
services~\cite{hidden-service-stats-techreport, hsstats-torproposal238,
owen2016empirical, tor-2015-01-001},
we avoid the need to store (even temporarily) onion addresses, since PSC
uses oblivious counters.
The results of our onion service address measurements are summarized in Table~\ref{tbl:PSC:onion:result}.

\paragraph{Unique Onion Service Addresses Published}
From a PSC measurement conducted between 2018-04-23 and 2018-04-24 during which
our mean combined HSDir publish weight was 2.75\%, we observed 3,900 (CI:
[3,769; 4,045]) unique v2 onion addresses in descriptors published to our HSDirs.
We extrapolate these results based on HSDir replication to infer that the total
number of unique onion service addresses published {\em for the entire Tor
network} is 70,826 (CI: [65,738; 76,350]).
Accordingly, our relays observed at least 4.93\% of Tor's unique onion addresses.
%
Using a different extrapolation method~\cite{hidden-service-stats-techreport},
the Tor Metrics Portal~\cite{tormetrics} estimates a total of 79 thousand unique
v2 onion services published in the Tor network at the time of our measurement.
The Tor metrics estimate does not include a confidence interval which we expect
to be significant since \textit{every} reporting relay adds noise (due to the lack of
secure aggregation of Tor metrics reports). Therefore, we expect that our
confidence intervals would overlap.

\paragraph{Unique Onion Service Addresses Fetched}
From a PSC measurement conducted between 2018-04-29 and 2018-04-30 during which
our mean combined HSDir fetch weight was 0.534\%,
we observed 2,401 (CI: [1,101; 3,718]) unique v2 onion addresses in descriptors
fetched from our HSDirs.
We extrapolate these results based on HSDir replication to infer that the total
number of unique onion service addresses fetched {\em for the entire Tor
network} is 74,900 (CI: [34,363; 696,255]).
%
By comparing the number of unique onion addresses published and fetched,
we estimate that between 45\% and 100\% of active onion services are
used by clients.
Note that the Tor Metrics Portal~\cite{tormetrics} does not estimate the number
of unique onion addresses fetched by clients, and its estimate of the number of
unique v2 onion addresses published occasionally varies significantly (\eg, it
increased by 40 thousand during our measurement).

\subsection{Onion Service Directory Usage}

\begin{table}[t]
  \centering
  \caption{Network-wide onion service descriptor statistics inferred from our HSDir PrivCount measurements.}
  \label{tbl:descstats}
  {\small
  \begin{tabular}{lll}
    \toprule
    {\bf Statistic} & {\bf Count} & {\bf 95\% CI} \\
    \midrule
    Fetched  	 & 134 million & [117; 150] million\\
    Succeeded    & 12.2 million & [10.6; 13.7] million \\ 
    Failed       & 121 million  & [103; 140] million \\
    Fail rate    & 1,400 failed/s & [1,192; 1,620] failed/s \\ 
    \midrule
    Public       & 56.8\% & [36.9; 83.6]\% \\
    Unknown      & 47.6\% & [28.8; 72.7]\% \\
    \bottomrule
  \end{tabular}
  }
\end{table}

While the number of unique onion \textit{addresses}
(\S\ref{sec:measurements:onion:address}) informs our understanding of how many
onion services exist, the number of \textit{descriptor} actions informs our
understanding of usage. Therefore, we use the PrivCount deployment described
in~\S\ref{sec:methodology:deploy} to safely measure the number of v2 onion
service descriptors that are 
\textit{fetched} from our
HSDirs. We also count the number of requests for descriptors that are not stored
in the DHT, typically because the service is inactive.
Our onion service descriptor measurement results are summarized in
Table~\ref{tbl:descstats}.

\paragraph{Onion Service Descriptors Fetched}
From a PrivCount measurement conducted between 2018-05-20 and 2018-05-21 during which
our mean combined HSDir fetch weight was 0.465\%, we infer that there were a
total of 134 million (CI: [117; 150] million) v2 descriptors fetched {\em in
the entire Tor network}.
Surprisingly, 90.9\% (CI: [87.8; 93.2]\%) of these fetches failed, \ie, the
corresponding descriptor was not present in the HSDir's cache or the request was
malformed. Our results indicate that there are \textit{at least} 103 million
failures per day, or about 1,192 failures per second. To be sure we did not
observe a network anomaly, we conducted this measurement multiple times and
observed consistent results.
We speculate that this large failure rate may be due to botnets or onion
site scanners with outdated onion address lists.
Tor's current statistics reporting infrastructure is
unable to collect HSDir lookup failures~\cite{tor-blog-onion-stats},
but we have demonstrated the feasibility of using the PrivCount protocol to
collect this statistic in a privacy-preserving manner.


From the measurement described above, we infer that there were a total of 12.2
million (CI: [10.6; 13.7] million) v2 descriptors successfully fetched {\em in
the entire Tor network}. For every successful descriptor fetch, we checked if
the address was available in the latest ahmia onion site search
index~\cite{ahmia}. We found that 56.8\% (CI: [36.9; 83.6]\%) of descriptors
fetched were present in the ahmia list at the time of our measurement, while
47.6\% (CI: [28.8; 72.7]\%) were not. Our results indicate that a majority of
successful onion service accesses are visits to
onion sites with publicly available addresses.

\subsection{Rendezvous Point Usage}
\begin{table}[t]
  \centering
  \caption{Network-wide rendezvous statistics inferred from our RP PrivCount measurements.}
  \label{tbl:rendstats}
  {\small
  \begin{tabular}{lll}
    \toprule
    {\bf Statistic} & {\bf Count} & {\bf 95\% CI} \\
    \midrule
    Total Circuits             & 366 million & [351; 380] million\\
    Succeeded  			 & 8.08\% & [3.47; 13.1]\%\\
    Failed: conn. closed & 4.37\% & [0.0; 9.23]\%\\
    Failed: circuit expired & 84.9\% & [77.0; 93.5]\%\\
    \midrule
    Cell payload     & 20.1 TiB & [15.2; 24.9] TiB \\ 
    Cell payload / second       & 2.04 Gbit/s & [1.55; 2.53] Gbit/s \\ 
    Cell payload / circuit & 730 KiB/circ. & [341; 2,070] KiB/circ. \\
    \bottomrule
  \end{tabular}
  }
\end{table}

Recall from \S\ref{sec:background:tor} that in order for a client to
communicate with an onion service, both parties build circuits to a relay called
a rendezvous point (RP), and that this rendezvous circuit carries all end-to-end
encrypted application payloads. While we cannot measure onion service usage by
counting streams (onion service streams are unobservable by our RPs because
data cells are end-to-end encrypted), we can infer rendezvous circuit activity
by counting the number of cells on the circuit. We instrument our RPs to report
the number of cells sent on v2~\cite{tor-rend-spec-v2} and
v3~\cite{tor-rend-spec-v3} rendezvous circuits, and use the
PrivCount deployment described in~\S\ref{sec:methodology:deploy} to safely
measure the approximate number of onion service rendezvous circuits and cells
relayed at our RPs.
Table~\ref{tbl:rendstats} summarizes the results of our RP measurements.

\paragraph{Rendezvous Circuit Activity}
From a PrivCount measurement conducted between 2018-05-22 and 2018-05-23 during
which our mean combined rendezvous weight was 0.88\%, we infer that there were a
total of 366~million (CI: [351; 380]~million) rendezvous circuits {\em in the
entire Tor network}. Note that since a successful rendezvous involves a client
and service circuit, each such rendezvous is counted as 2 circuits at the RP.
Surprisingly, only 8.08\% (CI: [3.47; 13.1]\%) of the circuits that we observed
succeeded and were active, \ie, they were used to transfer at least one cell containing an
application payload. We found that 4.37\% (CI: [0.0; 9.23]\%) of the observed
rendezvous circuits failed because the connection to the RP was closed before
the service completed the rendezvous protocol, and 84.9\% (CI: [77.0; 93.5]\%)
of the observed rendezvous circuits failed because the circuit expired
(timed-out) before the service completed the rendezvous protocol.

\paragraph{Rendezvous Circuit Data} 
From the same PrivCount measurement as above, we infer a total of 20.1~TiB (CI:
[15.2; 24.9]~TiB) of cell payload data on rendezvous circuits {\em in the entire
Tor network}. Our inference corresponds to a mean of 2.04~Gbit/s (CI: [1.55;
2.53]~Gbit/s) of payload data. By combining our circuit and cell observations,
we calculate that the mean amount of data per active Tor rendezvous circuit is
730~KiB (CI:[341; 2070]~KiB).

\section{Related Work \numpages{0.75}}
\label{sec:related}


Several efforts have attempted to improve our understanding of the
live Tor network.  Most notably, the Tor Project 
has allowed
researchers to access aggregate and longitudinal data about the Tor
network through its Tor Metrics Portal~\cite{tormetrics} since 2010,
with directory data spanning back to May 2004~\cite{tor-dirdata}.

\paragraph{Tor Metrics Portal}
The Metrics Portal indirectly estimates the number of Tor users by
counting the number of requests to the subset of Tor directory mirrors
that participate in statistics collection, and then extrapolating to
the entire network by dividing by the fraction of participating
directory mirrors.  This technique was originally proposed by Loesing
\etal~\cite{loesing2010measuring}.  In contrast with the Tor Metrics
Portal, our measurements are based on {\em direct} observations of
connecting users and do not depend on heuristics about how often
clients access directory mirrors.  (Indeed, as described in
\S\ref{sec:measurements:client}, our direct measurements do not
 align well with the Metric Portal's indirect estimates.)


\paragraph{Towards safe Tor measurements}
McCoy \etal~performed one of the first studies that attempted to
discern {\em how} users used Tor~\cite{tor-usage}.  There, the authors
performed packet capture at a small subset of Tor ingress and egress
points to determine the distributions of user locations and exit
traffic by port (i.e., by application).  Chaabane \etal~used similar
methods to examine BitTorrent behavior on
Tor~\cite{chaabane2010digging}.  Approaches such as these that collect
unobfuscated and potentially sensitive network traces have been
criticized on both ethical and legal
grounds~\cite{soghoian-cnet-2008,soghoian-wescr-2011}.

There has been an increasing effort in developing techniques to {\em
  safely} measure the Tor network.  Elahi et
al.~\cite{PrivExElahi2014} were the first to propose the use of
differential privacy to provide privacy-preserving aggregate data
about Tor.  Specifically, they propose two schemes known collectively
as PrivEx.  PrivCount~\cite{privcount-ccs2016}, which we use in our
measurements, extends the secret-sharing variant of PrivEx by
supporting repeatable measurement phases.
HisTor$\epsilon$~\cite{histore} also proposes the use of differential
privacy to safely measure statistics on anonymity networks, and adds
integrity protections by bounding the influence of malicious data
collectors (DCs).  However, compared to PrivCount, HisTor$\epsilon$ is
limited in the types of queries that it supports and lacks PrivCount's
mature implementation.  The private set-union cardinality (PSC)
protocol of Fenske \etal~\cite{psc-ccs2017} also provides
differentially private measurements.  We review differential privacy
in~\S\ref{sec:background:dp}.  Sections~\S\ref{sec:background:privcount}
and~\S\ref{sec:background:psc} describe PrivCount and PSC in more
detail.

\paragraph{Understanding Onion Services}
Goulet \etal~\cite{hidden-service-stats-techreport} describe the
benefits and privacy risks of statistics collection for onion
services.  They conclude with suggestions for privacy-preserving
statistics collection, including the use of differential privacy.
Their proposal partly inspired our work, in which we use differential
privacy and other techniques~\cite{psc-ccs2017} to conduct
measurements of onion sites.  Biryukov \etal demonstrate how the
design of onion services allows an attacker to gauge the popularity of
an onion service and potentially deanonymize
it~\cite{trawling-tor-hidden-services}.  However, their envisioned
attacks are not suitable for conducting privacy-preserving
measurements.
Recently, Jansen \etal show how to use PrivCount to safely measure the Tor
network to discover the popularity of onion
services~\cite{jansen2018insidejob}. Their work focused on traffic analysis
attacks and the popularity study of a single social networking onion service.
Finally, Owen and Savage perform empirical measurements
of Tor's onion services~\cite{owen2016empirical}.  We apply similar
techniques---operating Tor relays and observing HSDir lookups---but
also protect user privacy by using differentially private
techniques.

\section{Ethical and Safety\\Considerations \numpages{0.5}}
\label{sec:ethics}
Statistics collection in the setting of an anonymity network
inherently carries risk since the exposure of information could
degrade the privacy of the network's users and potentially subject
them to harm.  
Guiding our study
were the
four criteria for ethical network research established by the Menlo
Report~\cite{menlo-report}: we uphold the principle of {\em respect
  for persons} through the careful and principled application of data
protections, including the use of differentially private techniques and the
avoidance of collecting  personally identifiable information (e.g., IP
addresses).  Following the principle of {\em beneficience}, we balance
the (low) risk of harm with the potential benefits of the
research---namely, an increased understanding of the Tor network that
could inform research on improving the network's security and
performance.  Our techniques achieve the Menlo Report's notion of {\em
  justice}, since our statistics do not target a specific
subpopulation of Tor's users.  Finally, we achieve {\em respect for law
  and public interest} through transparency in our methods: we use
open source tools~\cite{privcount-ccs2016,psc-ccs2017} for statistics
collection and we submitted
our research plan (prior to its implementation) to several review
bodies. We presented it to the Tor Research Safety Board~\cite{trsb},
which concluded that our plan ``provides a plausible strategy for
safely measuring trends on the Tor network.''\footnote{The full text of
the TRSB's findings is available at
\url{https://security.cs.georgetown.edu/measurement-study/trsb-feedback.txt}.}
(Although some of the authors serve on
the TRSB, we were not involved in the TRSB's
deliberations of our research plan.)
Our measurement study was approved by the ethics board at the
University of New South Wales and certified as being exempt as
non-human subjects research by the Institutional Review Board (IRB) at
Georgetown University
Additionally,
aspects of this study were discussed with members of Georgetown's
Office of General Counsel, who raised no concerns.


\section{Conclusion and Discussion \numpages{0.25}}
\label{sec:conclusion}

This paper presents the most comprehensive privacy-preserving
measurement study of the live Tor network to date.  Our findings
confirm that Tor is used predominantly for web browsing and that Tor
users tend to visit the same popular (\ie, top Alexa) sites as do
ordinary (and non-anonymous) Internet users. We observe a surprisingly
large number of Tor connections and unique client IP addresses,
suggesting that the heuristically-derived estimates from the Tor
Metrics Portal are significantly underreporting Tor's userbase.  We
find that the
network's clients are highly distributed, connecting from more than
200 countries and nearly 12,000 ASes.  These clients construct more
than 1.2 billion anonymous circuits per day, carrying approximately
517 TiB of data (6.1GiB/s).  Unexpectedly, our
measurements show that more than 90\% of onion service descriptor
fetches fail, suggesting the presence of botnets on Tor or aggressive
onion site crawlers with outdated address lists.  We report 20 TiB of
data transferred daily over circuits to onion services, representing
roughly 3.9\% of all Tor traffic.

Our study does not attempt to distinguish
between human-driven and automated activity on the Tor network.  An
intriguing open problem is the construction of techniques to identify
when measurements are heavily influenced by automated activities (for
example, when swarms of infected hosts belonging to a botnet connect
to Tor).  Indeed, the enormously high failure rate of onion service
descriptor fetches strongly suggests that automated behavior
likely does occur on Tor.  Our hope is that measurement studies such
as this will help inform developers and security researchers of
unexpected automated activities---a necessary prerequisite to
understanding and potentially defending against malicious automated
processes.

\section*{Acknowledements}

This work has been partially supported by the Office of Naval Research, the
National Science Foundation under grant number CNS-1527401, and
the Department of Homeland Security Science and Technology Directorate, Homeland
Security Advanced Research Projects Agency, Cyber Security Division under
agreement number FTCY1500057. The views expressed in this work are strictly
those of the authors and do not necessarily reflect the official policy or
position of any employer or funding agency.

\balance
\bibliographystyle{plain}
\bibliography{references,micah-long}
\end{document}